\documentclass{article}

\usepackage{arxiv}
\usepackage[colorinlistoftodos]{todonotes}
\usepackage{amstext}
\usepackage{xspace}
\usepackage{booktabs}
\usepackage{graphicx}
\usepackage{hyperref}
\usepackage{cleveref}
\usepackage{multirow}
\usepackage{subcaption}
\usepackage{colortbl}
\usepackage{natbib}
\Crefformat{figure}{#2Fig.~#1#3}
\Crefmultiformat{figure}{Figs.~#2#1#3}{ and~#2#1#3}{, #2#1#3}{ and~#2#1#3}

\newcommand{\name}{\textsc{Atropos}\xspace}
\newcommand{\lc}[1]{$LC(#1)$\xspace}
\AtBeginDocument{%
  }

\begin{document}

\title{\name: Improving Cost-Benefit Trade-off of LLM-based Agents under Self-Consistency with Early Termination and Model Hotswap}

\author{
Naryeong Kim\\
KAIST\\
Daejeon, Republic of Korea\\
\texttt{naryeong.kim@kaist.ac.kr} \\
\And
Shin Yoo\\
KAIST\\
Daejeon, Republic of Korea\\
\texttt{shin.yoo@kaist.ac.kr} \\
}

\maketitle

\begin{abstract}
Open-weight Small Language Models(SLMs) can provide faster local inference at lower financial cost, but may not achieve the same performance level as commercial Large Language Models (LLMs) that are orders of magnitudes larger. Consequently, many of the latest applications of LLMs, such as software engineering agents, tend to be evaluated on larger models only, leaving the issue of improving the cost-benefit trade-off of such applications neglected. This paper proposes \name\footnote{Atropos, the Greek goddess of fate, cuts the thread of human life. The name suggests that our technique performs early-termination of LLM inferences.}, a predictive early-termination analysis and hotswap technique that aims to improve the cost-benefit trade-off for LLM-based agents that use self-consistency. The core component of \name is a predictive model based on structural properties of LLM inferences: after merging multiple agentic inference paths into a graph representation, \name uses Graph Convolutional Network (GCN) to predict whether an ongoing inference will eventually succeed or not. If an agentic task instance running on the source LLM is predicted to fail, \name subsequently performs hotswapping, i.e., migrating the on-going inference context onto the more capable target LLM: this is feasible because LLM contexts are stateless. An empirical evaluation of \name using three recent LLM-based agents shows that \name can predict early termination of eventually failing inferences with the accuracy of 0.85 at the midpoint of the inference. Hotswapping LLMs for such inferences can convert up to 27.57\% of them to be successful. Consequently, \name achieves 74.35\% of the performance of closed LLMs with as low as only 23.9\% of the cost. 
\end{abstract}


\section{Introduction}
\label{sec:introduction}

Agentic system design has been rapidly adopted to automate software engineering tasks that cannot be easily tackled by engineering a single prompt. Agents designed for specific software engineering tasks such as fault localization~\cite{kang2024autofl} and program repair~\cite{zhang2024acr,bouzenia2025ra} are rapidly advancing thanks to the improving performance of the underlying Large Language Models (LLMs) as well as large, realistic benchmarks~\cite{jimenez2024swe}. These agentic applications typically depend on, and are evaluated with, large and powerful proprietary LLMs, at significant financial, communication, and computational cost. In contrast, open-weight Small Language Models (SLMs) are often sufficiently lightweight to be deployed locally with low latency, reducing the cost of LLM queries. However, SLMs are typically less powerful than their proprietary alternatives, resulting in the emphasis on proprietary models. 

We note that the cost-benefit trade-off of agentic systems remains largely unexplored. Given that agentic applications are increasingly adopting longer and more complicated iterative reasoning, we expect the inference cost to continue to rise. The use of self-consistency~\cite{wang2023sc}, i.e., taking multiple inference samples to improve the performance~\cite{Ahmed2023aa}, can only amplify the cost increase. SLMs may not be the ideal cost reduction solution for more expensive agentic inferences, since the lower performance of SLMs will result in more failures than agents driven by proprietary LLMs. More importantly, such failures will only be detected once we have spent the computational resources on SLMs, highlighting the need for more effective optimization of the cost-benefit trade-off.

This paper introduces \name, a predictive early-termination and LLM hotswapping technique designed to improve the cost-benefit trade-offs of LLM-based agent applications. \name operates by formulating ongoing agentic inferences as Semantic Flow Graphs (SFGs)~\cite{Yoo2025br} and predicting, as early as possible, whether the current inference is likely to succeed or fail if the agent is allowed to continue on the current LLM. \name uses a Graph Convolutional Network (GCN) to predict the eventual success of the current inference that is currently running on a local SLM. If a failure is predicted, \name performs \emph{hotswapping} by migrating the ongoing context to a more capable remote LLM. This is feasible because LLM queries are stateless, allowing \name to continue an ongoing inference on a new LLM simply by replaying the inference up to that point.

We evaluate \name across three state-of-the-art agents, AutoFL~\cite{kang2024autofl}, AutoCodeRover~\cite{zhang2024acr}, and RepairAgent~\cite{bouzenia2025ra}, under self-consistency with ten samples: we aim to predict whether the current inference is leading to  accurate fault localization for AutoFL and AutoCodeRover, and correct patch generation for RepairAgent. The empirical evaluation shows that \name can correctly predict early termination of inferences unlikely to succeed with the accuracy of up to 0.85, and AUROC of 0.85 at the midpoint of the inference. Performing LLM hotswap for inferences predicted to be unsuccessful converts an up to 27.57\% of them to be successful. Consequently, \name achieves success in 74.35\% of successful inferences performed using proprietary LLMs, using 23.90\% of total monetary cost. 

The technical contribution of this paper is as follows:

\begin{itemize}
  \item We show that accurate prediction of eventual success of ongoing LLM inferences is possible: \name achieves accuracy of up to 85.4\% and AUROC of 85.45\% when predicting whether an LLM inference will result in an incorrect answer at the midpoint of the inference.

  \item We show that hotswapping LLMs can salvage some of the inferences that were predicted to fail on SLMs. By migrating the ongoing inference contexts onto a more capable LLM, \name can lead up to 27.57\% of inferences that were going to fail if they were allowed to continue on SLMs. 

  \item We empirically show that, using \name, it is possible to improve the cost-benefit trade-offs of LLM-based agent applications. \name reduces the monetary cost by up to 76.1\% compared to proprietary LLMs, while still retaining 74.35\% of their performance.

  \item We compare various configurations of the early termination prediction using Semantic Flow Graphs. Depending on the nature of the task we apply \name to, how to represent the LLM inferences as an SFG can vary. Our results suggest that \name can be generalized to other agents, as it retains the early termination prediction performance even when SFG construction involves embeddings of unstructured texts as part of agent actions (such as code patches and natural language descriptions).

\end{itemize}

The rest of the paper is organized as follows. Section~\ref{sec:preliminaries} provides preliminary information, on top of which Section~\ref{sec:method} presents \name in detail. Section~\ref{sec:experimental_setup} describes the settings of our empirical evaluation, whose results are presented in Section~\ref{sec:results}. Section~\ref{sec:threats_to_validity} discusses threats to validity, and Section~\ref{sec:related_work} presents related work. Section~\ref{sec:conclusion} concludes.

\section{Preliminaries}
\label{sec:preliminaries}

Here we present some background information, as well as the agents we study to evaluate \name.

\subsection{Prompt Engineering}
\label{sec:prompt_engineering}

Various prompt engineering techniques have been proposed to exploit the emergent capabilities of LLMs~\cite{sahoo2024pe}. In this paper, we focus on ReAct~\cite{yao2023react}, which enables tool usage of LLM agents, and self-consistency~\cite{Wang2023aa}, which improves the performance of LLM agents using multiple samples of the same query.

\subsubsection{ReAct}
\label{sec:react}

ReAct~\cite{yao2023react} generates reasoning traces and task-specific actions in an interleaved manner. It allows models to decompose tasks and dynamically update plans based on environmental feedback. For example, to answer a multi-step question, an agent reasons about missing information, invokes necessary APIs to obtain such information, and utilizes the observations of the results to guide subsequent actions. ReAct has been widely adopted by software engineering agents~\cite{kang2024autofl, zhang2024acr, bouzenia2025ra, rondon2025passerine}. These agents iteratively invoke external tools to solve tasks, producing trajectories in which each step typically corresponds to an invocation of an external tool.

\subsubsection{Self-Consistency}
\label{sec:self-consistency}

Self-consistency~\cite{wang2023sc} is a prompt engineering technique that asks the same question multiple times and aggregates the results to make a final inference. The underlying assumption is that the correct answer is likely to be the most consistent one across multiple samples. Since there is typically a unique correct solution, correct reasoning paths inevitably converge to the same result. Conversely, given the vast space of possible incorrect answers, each incorrect reasoning is likely to produce its own different answer.

Since applying self-consistency requires multiple executions, it generates a set of candidate answers. In particular, when integrated with multi-step reasoning frameworks like ReAct~\cite{yao2023react}, this approach yields multiple execution trajectories corresponding to each reasoning path. Consequently, while self-consistency can improve the performance of LLM-based agents, it also incurs high computational costs due to the requirement of multiple executions that include tool invocations.

\subsection {Agents}
\label{sec:agents}

With the advances in reasoning capabilities of LLMs, many LLM-based agents have been proposed for software engineering tasks~\cite{yang2024swe, wang2025openhands, zhang2024codeagent}. This paper focuses on three agents whose primary actions are iterative tool invocations based on ReAct: AutoFL~\cite{kang2024autofl}, AutoCodeRover~\cite{zhang2024acr}, and RepairAgent~\cite{bouzenia2025ra}. Here we briefly introduce each agent. 

\subsubsection{AutoFL}
\label{sec:autofl}

AutoFL~\cite{kang2024autofl} performs the Fault Localization (FL) task at the method level: given the source code of a failing test case, it identifies buggy locations based on test execution results, coverage data, and source code. In particular, AutoFL can navigate the code repository to gather necessary information and narrows down the search space of suspicious code, using external functions, i.e., tools. Initially, the LLM is provided with information regarding failing test cases and descriptions of the following four functions: \verb|get_failing_test_covered_classes| returns a list of classes covered by the failing test, \verb|get_failing_test_covered_method_for_class| returns a list of methods (belonging to a specified class) covered by the failing test, \verb|get_code_snippet| returns the source code for a specified method, and \verb|get_comment| returns the docstring for a specified method. AutoFL is instructed to start with the list of classes covered by the failing test, and subsequently navigates the code repository using the tools. 

To enhance performance, AutoFL adopts Self-Consistency~\cite{wang2023sc}. It repeats the localization process multiple times and selects the final answer through majority voting. Further, AutoFL introduces the concept of confidence, defined as the voting score of the final answer, to quantify the level of certainty in its predictions.

\subsubsection{AutoCodeRover}
\label{sec:autocoderover}

AutoCodeRover (ACR)~\cite{zhang2024acr} is an Automated Program Repair (APR) agent: it aims to resolve issues in software projects with a range of tools including code search, Spectrum Based Fault Localization (SBFL)~\cite{Wong:2016aa}, and linter. Using Abstract Syntax Tree (AST) based code search, ACR iteratively gathers bug-related contexts and eventually generates patches. 

AutoCodeRover divides the workflow into FL and patch generation. In this work, we only use the FL stage of ACR because our proposed technique depends on self-consistency. Answers produced in the FL stage are discrete code identifiers (e.g., method signatures), which in turn facilitates consensus aggregation through majority voting, whereas such voting is difficult for patches because semantically identical patches may still be syntactically different from each other. To study the FL performance of ACR, we use a configuration without SBFL support: ACR is equipped with eight functions, including  \verb|search_class|, \verb|search_method_in_class|, and \verb|search_code_in_file|. The trajectories of AutoCodeRover capture the exploration strategy to navigate the given repository.

\subsubsection{RepairAgent}
\label{sec:repairagent}

RepairAgent~\cite{bouzenia2025ra} is another agent designed for APR. It performs program repair by performing three distinct subtasks: \emph{understanding the bug}, \emph{collecting information to fix the bug}, and \emph{trying to fix the bug}. Each subtask corresponds to an internal state that provides different sets of tools. The initial state for understanding the bug  employs functions such as \verb|run_tests| and \verb|extract_tests| to gather context regarding failing test cases and potential buggy locations. The second state for collecting information provides functions such as \verb|search_codebase| and \verb|get_classes_and_methods|. The final state of patch generation involves writing and executing candidate patches. In total, RepairAgent uses a suite of 12 functions, including tools whose function is to transition to specific target states. Consequently, the trajectory of RepairAgent captures its entire repair workflow, including the sequence of functions and state transitions.

Similar to AutoCodeRover, the final output of RepairAgent is code patches, making it difficult to aggregate results from multiple samples for self-consistency through voting. However, the state-based architecture of RepairAgent does not allow easy separation of FL stage. As a result, we use RepairAgent in its entirety and report the number of executions that produce plausible patches.

\subsection{Semantic Flow Graphs}
\label{sec:semantic_flow}

To analyze trajectories produced by different agents, we need to adopt a common data structure. Semantic Flow Graph (SFG) is a recently introduced representation for ML-based systems~\cite{Yoo2025br}: it combines traditional control flow and the semantic information that drives Machine Learning (ML) components, such as LLMs, in a graph format, so that the behavior of ML components (in our case, the decisions about tool invocation made by the LLM) can be considered in the context of the execution of the entire software system (in our case, the entire trajectory of an LLM-based agent). 

We adopt SFG as the graph representation of agent trajectories, as the agents studied in this work are typical examples of such ML-based software systems --  they consist of both traditional software (i.e., the agent framework as well as implementation of individual tools) and ML components (i.e., LLMs that drive the agent inference). Specifically, we represent each tool invocation with nodes in SFG, and the invocation sequence with edges in SFG. See Section~\ref{sec:graph_construction} for more details.

\begin{figure}[ht]
  \centering
  \includegraphics[width=0.8\textwidth]{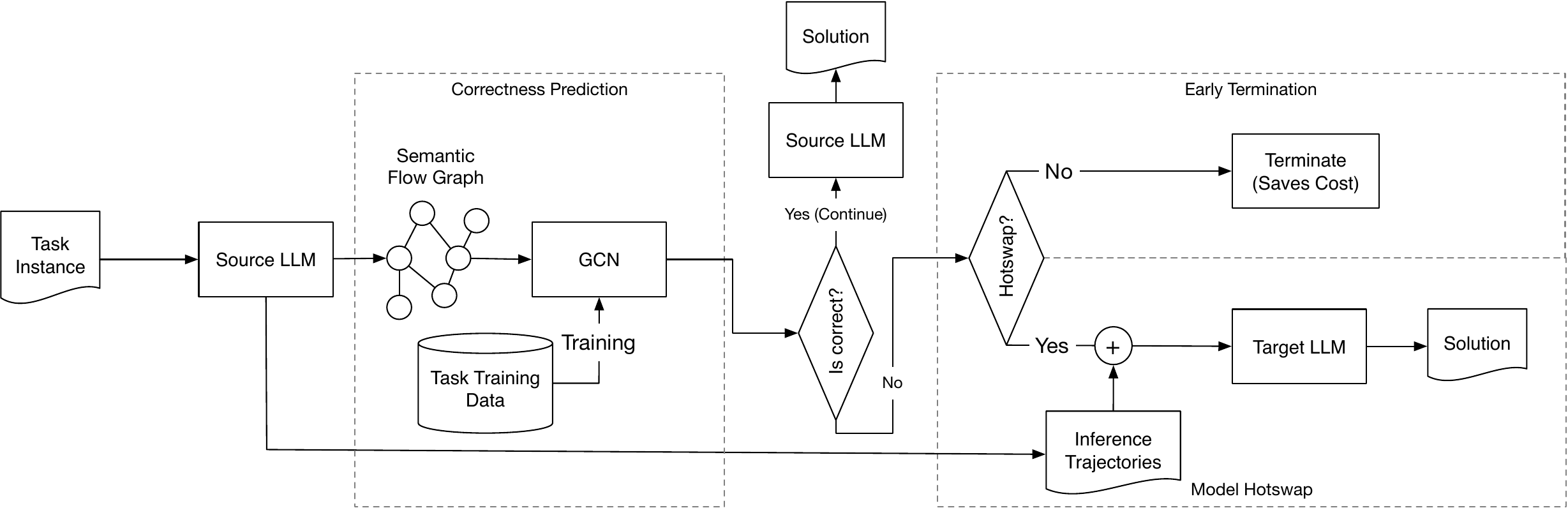}
  \caption{Overall workflow of \name\label{fig:overview}}
  \end{figure}

\section {\name: Early Termination Prediction and Model Hotswap}
\label{sec:method}

\name is composed of two major parts: early termination prediction, and model hotswapping (\Cref{fig:overview}). Given a new instance of an agent task, \name first runs the agent on a local SLM, taking multiple samples for self-consistency. During the inference, \name constructs an SFG of inferences performed up to that point, and uses a Graph Convolution Model trained to predict whether an SFG representing partial inferences would result in a successful solution or not. Depending on the prediction, the user can terminate inferences unlikely to succeed, thereby saving the inference cost. However, if the task cannot be abandoned, the user may choose to continue after hotswapping the model with a more capable one. For hotswapping, \name migrates the LLM context of the original source LLM to the target LLM, simply by replaying the inferences up to that point in the context of the target LLM. The following subsections explain these steps in more detail. 

\subsection{Graph Construction}
\label{sec:graph_construction}

To predict the correctness of the consensus of multiple samples under self-consistency~\cite{wang2023sc}, we adopt the existing work~\cite{kim2025lachesis} and construct an SFG that aggregates multiple inference trajectories~\cite{Yoo2025br}. Existing work, Lachesis~\cite{kim2025lachesis}, was proposed specifically for AutoFL~\cite{kang2024autofl} and only predicted the correctness label for trajectories of inferences that have been completed. \name extends this framework to enable prediction with partial inferences for early termination, also to support more generic node representations for other agents.

\begin{figure}[ht]
  \centering
  \includegraphics[width=0.7\textwidth]{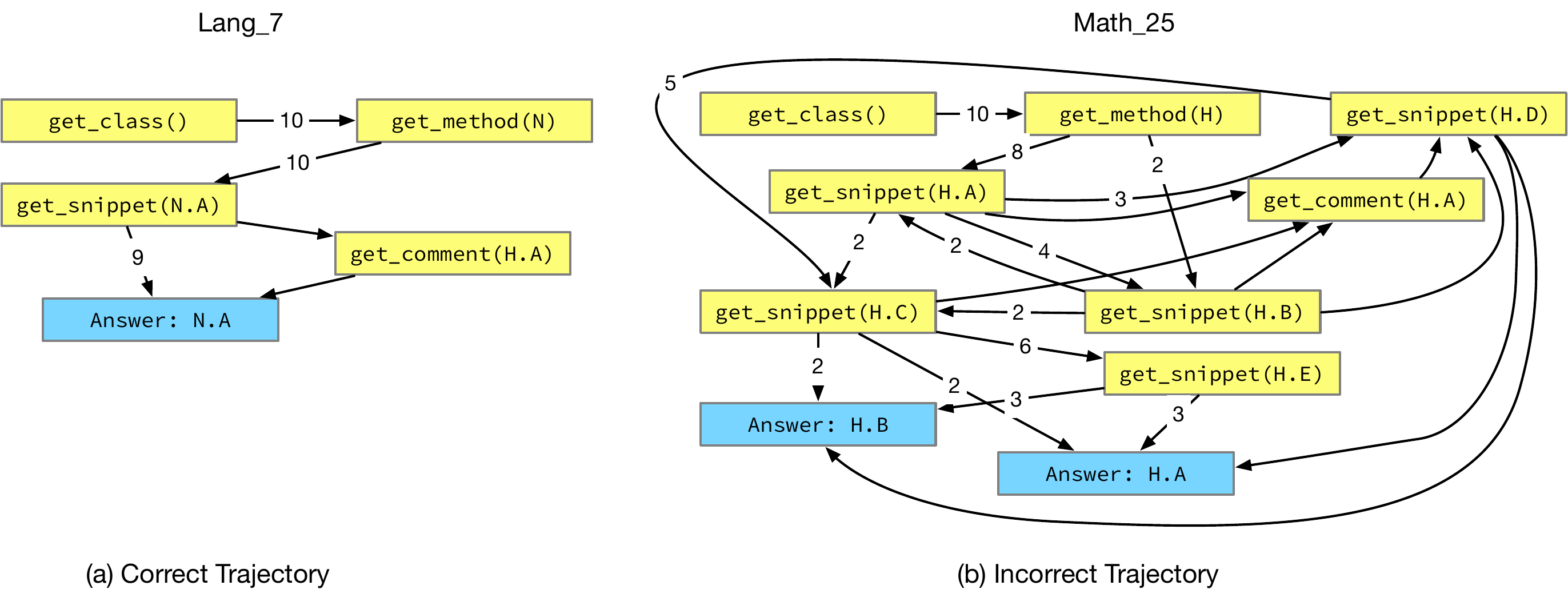}
  \caption{Examples of SFG taken from inferences of AutoFL. Edge weights of 1 are not explicitly labeled.\label{fig:sfg}}
  
 \end{figure}

We construct SFGs so that nodes represent unique reasoning steps, and weighted directed edges denote the sequential transitions between them. This structure is designed to encapsulate the structural properties of multiple trajectories. Specifically, the graph not only preserves the sequential flow of individual trajectories but also highlights reasoning sequences that recur frequently across multiple executions, which are represented by edges with higher weights. Figure~\ref{fig:sfg} shows example SFGs constructed from correct and incorrect inference trajectories of AutoFL for two different faults in Defects4J.

\subsubsection{Node and Edge Representation}
\label{sec:representation}

To facilitate prediction using GCN, nodes in the SFG need to be embedded into vector forms. In turn, the key requirement for SFG node embedding is that the same reasoning step should share the same embedding, so that the characteristics of reasoning paths are preserved in the graph. For AutoFL and AutoCodeRover, comparing identities of reasoning steps is straightforward, as their function calls utilize deterministic code signatures like class and method names as arguments. Such grouping of reasoning steps results in SFGs like ones shown in \Cref{fig:sfg} (note that arguments such as \verb|N.A| and \verb|H.B| simply stand for different identifiers). Once grouped, each node needs to represent the specific tool invocation of that reasoning step as well as the arguments of that invocation. For AutoFL and ACR, we use one-hot encoding to represent tools, and FastText~\cite{bojanowski2016fasttext} embedding with 300 dimensions to represent all arguments. Given a set of tools, $F$, we use one-hot vector of length $|F| + 1$ to accommodate $|F|$ tools as well as abnormal tool invocation: the final dimension of each node for AutoFL and ACR therefore is $|F| + 1 + 300$. ACR may invoke multiple tools in a single reasoning step, for which we use multi-hot embedding of the same dimension $|F| + 1$. Once we embed all nodes, edges simply represent the sequences of reasoning steps in the given self-consistency trajectory samples: each edge is weighted by the frequency of two connected reasoning steps appearing consecutively across the trajectories.

The construction of SFG is more complicated for RepairAgent: tools like \verb|express_hypothesis| and \verb|write_fix| accept unstructured arguments, such as natural language descriptions and code snippets. Constructing graph based on exact string matching, as employed in AutoFL and AutoCodeRover, becomes infeasible. Instead, we employ a clustering approach to group semantically similar reasoning steps. In the resulting graph, each node represents a cluster of reasoning steps rather than a unique, identical function call. Similar to AutoFL and ACR, each reasoning step is encoded by concatenating a one-hot function vector and a 300-dimensional FastText~\cite{bojanowski2016fasttext} argument vector. We then employ an incremental clustering strategy based on cosine similarity. Each incoming reasoning step is assigned to the cluster whose centroid yields the highest similarity exceeding a predefined \emph{assignment threshold}. Cluster centroids are updated dynamically as more steps are added. Finally, to minimize redundancy, we merge clusters whose centroids exhibit similarity surpassing a \emph{merge threshold}, and update their centroids. We empirically set both thresholds to 0.99 in this study. Each node corresponds to one of the final clusters, and is represented by the cluster centroid. Edges are determined by tracing the sequence of reasoning steps within the original trajectories. For every transition from one step to the next, we add or increment a weighted directed edge between their respective clusters. Notably, this definition allows for self-loops, which are generated when two consecutive steps belong to the same cluster.

By clustering reasoning steps via FastText~\cite{bojanowski2016fasttext} embeddings, \name aims to construct graph representations of inferences that are robust to syntactic variations: we expect the clustered trajectories to share similar semantic information, similar to the concept of semantic entropy~\cite{farquhar2024se}.

\subsubsection{Trajectory Truncation}
\label{sec:truncation}

Unlike existing work that predicted the correctness label for completed inference trajectory for AutoFL~\cite{kim2025lachesis}, \name aims to predict whether the set of ongoing inferences will eventually succeed or not, so that it can intervene with early termination. For this, we need to construct training datasets that represent partial inferences. We do this by running full inferences and subsequently truncating the inference trajectories.

\begin{figure}[ht]
  \includegraphics*[width=\textwidth]{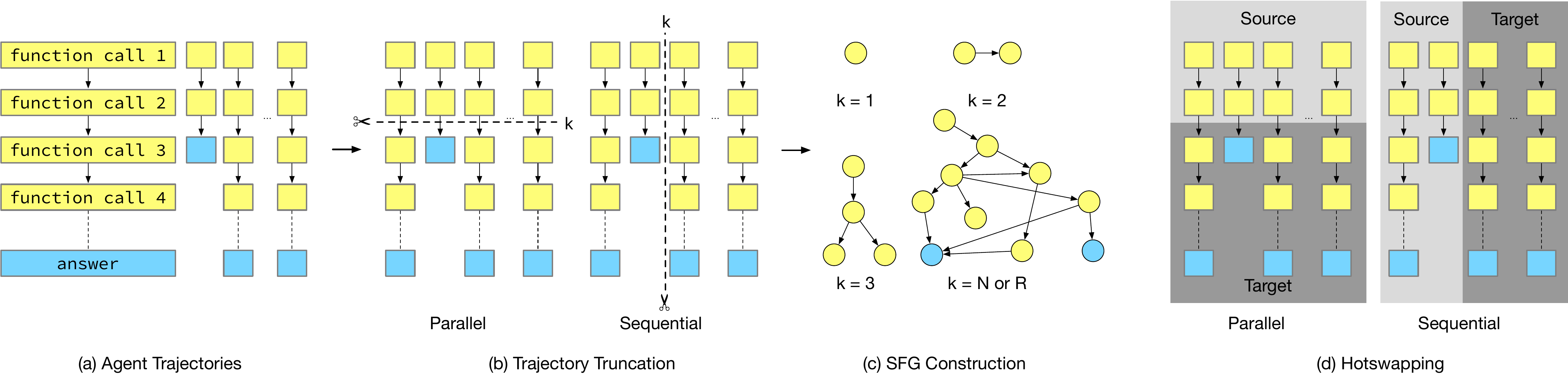}
  \caption{Trajectory Truncation, SFG Construction, and Model Hotswap
  \label{fig:truncation}}
\end{figure}

Let us assume that each agent operates within an interaction budget of $N$, i.e., it can make at maximum $N$ function invocations. Further, let us also assume that we apply self-consistency with $R$ queries. In total, this yields $R$ trajectories, each with a maximum length of $N+1$ (the additional step represents the final answer generation step used by AutoFL and ACR; RepairAgent does not have a separate output generation step, and therefore its trajectories are of length $N$). To assess prediction accuracy using comprehensive structural information, we first utilize $R$ complete trajectories, each with maximum length of $N+1$, as illustrated in \Cref{fig:truncation} (a). 

\name supports two different types of early termination. The parallel approach assumes that the user is performing $R$ inferences in parallel, but step by step: early termination for parallel approach means terminating $R$ partial inferences. In contrast, the sequential approach assumes that the user is performing $R$ inferences by completing each inference one by one: early termination for sequential approach means terminating after $k$ complete inferences ($k \leq R$). To construct partial trajectory datasets that correspond to each scenario, we truncate the complete trajectories in two different ways, as shown in \Cref{fig:truncation} (b): 

\begin{itemize}
\item \textbf{Parallel Truncation:} we truncate all $R$ trajectories after the first $k$ reasoning steps. Any trajectories whose length is shorter than $k$ remain intact. Consequently, the resulting graph contains at most $k \times R$ reasoning steps.
\item \textbf{Sequential Truncation:} we retain the first $k$ fully completed trajectories and discard the remaining $R - k$ trajectories. Each of these $k$ can use up to $N$ tool invocations, resulting in graphs that contain at most $(N+1) \times k$ reasoning steps.
\end{itemize}

For each studied agent, we construct multiple truncated trajectory datasets with varying $k$, with which we train the Graph Convolution Network that predicts the correctness of the inferences.

\subsection{Correctness Prediction for Early Termination}

We now train a predictive model using the truncated inference trajectory datasets.

\subsubsection{Labeling Criteria}

The correctness prediction is a binary classification problem: we assigned a label of $y=1$ for failure cases and $y=0$ for success cases. The specific criteria for each agent are defined as follows:

\begin{itemize}
\item\textbf{AutoFL}
Since AutoFL~\cite{kang2024autofl} performs method-level fault localization, we apply majority voting to select the top-$n$ suspicious methods. If the ground truth buggy method is not present among the top-$n$ candidates (acc@$n$), the SFG that corresponds to the final outcome is labeled as a failure ($y=1$); otherwise, it was labeled as a success ($y=0$).

\item\textbf{AutoCodeRover} 
While AutoCodeRover~\cite{zhang2024acr} outputs localization results at varying granularity (file, class, or method), we adopt majority voting on the output signatures without distinguishing between these levels. We utilize the developer-provided gold patch from SWE-bench~\cite{jimenez2024swe} as the ground truth. If any of the top-$n$ voted signatures does not correspond to code regions modified by the gold patch (acc@$n$), the SFG that correspond to the final outcome is labeled as a failure ($y=1$); otherwise, it is labeled as a success ($y=0)$.

\item\textbf{RepairAgent} For the APR task of RepairAgent, integrating generated patches via majority voting is infeasible due to the diversity of code. Instead, we label the inferences based on the number of executions generating plausible patches out of $R$ runs ($N_{ppe} \ge n$). If the number of executions does not exceed a predefined threshold $n$, we labeled it as a failure ($y=1$); otherwise, it is labeled as a success ($y=0$).
\end{itemize}

For the sake of brevity, we refer to the configuration of Labeling Criteria $acc@1$ (for AutoFL and ACR) and $N_{ppe} \ge 1$ (for RepairAgent) as \lc{1}. Similarly, we denote the criteria corresponding to thresholds of $n=3$ and $n=5$ for $acc@n$ and $N_{ppe} \ge n$ as \lc{3} and \lc{5}, respectively. Unless specified otherwise, we primarily report results based on \lc{1} to evaluate the performance for consistency. However, note that, for RepairAgent, \lc{5} is a more difficult criterion than \lc{1}, as it requires more executions that produce plausible patches.

\subsubsection{Prediction Model}
\name employs Graph Convolutional Network (GCN)~\cite{Kipf2016gcn} to perform graph-level binary classification. We use three GCN layers followed by a readout layer. Each of the first two GCN layers are connected to a ReLU activation layer followed by a dropout layer. The output from the final GCN layer is aggregated via global mean pooling into a graph embedding, which is then processed by a linear layer to generate the final prediction. If the prediction is sufficiently accurate, it can be used as a signal to terminate inferences that are unlikely to succeed early on, thereby saving inference cost.

\subsection{Hotswapping}

While early termination itself can save inference cost by not wasting tokens for inferences that would eventually fail, resolving such unpromising tasks may be critical for the user. A natural solution would be to try to resolve the same task with a more powerful LLM: in the literature, this is referred to as model cascading~\cite{ramirezOptimisingCallsLarge2024a}. However, cascading assumes that the original inference has finished. Even if the subsequent queries to a more capable LLM resolve the task, the user has wasted the inference cost incurred by the initial model. \name aims to overcome this issue by hotswapping, i.e., by \emph{continuing} the inauspicious inference using a more capable LLM.

Let us assume that we have two models: a cost-effective \emph{source model} and a more powerful \emph{target model}. Initially, the inference starts with the source model. During the process, the GCN model predicts the correctness of the ongoing inference. If the inference is predicted to fail, we switch to the target model while maintaining the current context. Similar to the truncation strategy in \Cref{sec:truncation}, hotswapping can be conducted in two ways: parallel and sequential.

\subsubsection{Parallel Hotswap}

We perform correctness prediction at a specific step $k$ for all $R$ ongoing inferences. When inferences observed up to step $k$ are predicted to fail collectively, we trigger hotswapping as depicted in the left side of \Cref{fig:truncation} (e). Specifically, we migrate the inference contexts of the first $k-1$ steps from all $R$ inferences performed on the source model, to the target model. Subsequently, we continue to generate the trajectories from step $k$ onward on the target model.

We note that hotswapping is only applied to active trajectories that have not terminated on the source model by step $k$. We retain the results of early-terminated trajectories because attempting to modify a completed trajectory would necessitate a full re-execution. Consequently, we achieve seamless context migration without incurring the overhead of re-executing already completed tasks. However, this also implies that the higher the value of $k$ is, the lower the hotswap performance becomes, as \name has to work with more reasoning steps from the weaker source model.

\subsubsection{Sequential Hotswap}

A sequential hotswap is initiated for the remaining executions, when the first $k$ complete trajectories generated by the source model are predicted to fail, as shown in the right side of \Cref{fig:truncation} (b). For instance, assuming a total of $R=10$ executions, if the model predicts a failure at $k=5$, the final consensus result is aggregated using the first four trajectories from the source model, and the remaining six trajectories are generated by the target model. A sequential hotswap therefore does not involve context migration; it can be considered as switching to a more capable ensemble of LLMs on the fly.

\section{Experimental Setup}
\label{sec:experimental_setup}

We present our experimental setup in this section.

\subsection{Research Questions}
We designed our experiments to evaluate the effectiveness and efficiency of \name by addressing the following research questions:

\begin{itemize}
  \item \textbf{RQ1. Correctness Prediction Accuracy:} Does the SFG representation capture sufficient information indicative of failing agent inferences so that GCN can accurately predict the correctness of trajectories? We construct SFGs using full trajectories without truncation, train the GCN model, and evaluate the predictive accuracy of correctness. We answer RQ1 by comparing accuracy, AUROC, AUPR and FPR@95 against baselines.

  \item \textbf{RQ2. Effectiveness of Early Termination:} How early can \name reliably predict eventual task failures for early termination? We generate \name graphs using both parallel and sequential truncated trajectories for each step $k$ and evaluate prediction performance of \name. Further, we also investigate the trade-off between cost savings and agent performance to identify preferable termination points.

  \item \textbf{RQ3. Hotswap Effectiveness:} How does the model hotswapping technique improve the cost-benefit trade-off compared to single-model baselines? We answer RQ3 by analyzing the performance and cost implications of both parallel and sequential hotswapping across varying hotswap points $k$.

  \item \textbf{RQ4. Ablation Study:} Which components of SFG representation contribute the most to the predictive power of \name? We answer RQ4 by varying node representation schemes for SFGs of full trajectories and comparing the predictive performance of GCN for the correctness of inferences using accuracy and AUROC metrics.
\end{itemize}

\begin{table}[t]
  \caption{Configuration for agents, benchmarks, and models ($N$: \# of Steps, $R$: Self-consistency Sample Size)}
  \label{tab:setup}
  \centering
  \setlength{\tabcolsep}{8pt}
  \scalebox{0.9}{
  \begin{tabular}{llccll}
    \toprule
    \textbf{Agent} & \textbf{Benchmark (Size)} & \boldmath{$N$} & \boldmath{$R$} & \textbf{Source Model} & \textbf{Target Model} \\ \midrule
    AutoFL & D4J (353) + BIP (500) & 10 & 10 & Llama-3-8B & GPT-4o \\
    AutoCodeRover & SWE-bench (1000) & 15 & 5 & Mixtral-8x7B & GPT-4 \\
    RepairAgent & D4J (605) & 40 & 10 & GPT-3.5-turbo & GPT-4o \\ \bottomrule
  \end{tabular}}
\end{table}

\subsection{Agents, Benchmarks, and Models}

We evaluate \name using three software engineering agents: AutoFL~\cite{kang2024autofl}, AutoCodeRover~\cite{zhang2024acr}, and RepairAgent~\cite{bouzenia2025ra}, as described in Section~\ref{sec:preliminaries}. \Cref{tab:setup} summarizes the detailed configuration for each agent, including the datasets, interaction budgets ($N$), self-consistency sample size ($R$), and the specific language models used as source and target. 

For the model configuration, our intention is to pair a cost-effective \emph{source} model for initial execution with a more capable \emph{target} model for hotswapping. For AutoFL and AutoCodeRover, we employ open-weight SLMs, Llama-3-8B~\cite{grattafiori2024llama} and Mixtral-8x7B~\cite{jiang2024mixtral} as source models, and proprietary LLMs, GPT-4o and GPT-4~\cite{openai2024gpt4},\footnote{ACR is evaluated using GPT-4 in the original paper~\cite{zhang2024acr}.} respectively.\footnote{Llama-3-8B results in poor FL performance with ACR, which is why we switched to Mixtral-8x7B.} In the case of RepairAgent, generating valid patches for the APR task remains challenging for current open-weight SLMs. To ensure a balanced dataset for training our GCN model, we adopt GPT-3.5-turbo as the source model for RepairAgent, treating it as the lower-cost alternative to GPT-4o.

\subsection{Baselines \& Evaluation Metrics}
To the best of our knowledge, there is no existing baseline specifically designed to predict the correctness of partial inferences under self-consistency. Therefore, we compare \name against the following baselines to assess its effectiveness:

\begin{itemize}
  \item \textbf{Majority Class Baseline:} To present the inherent class balance within the dataset, we report the accuracy obtained by a baseline that classifies all instances as the majority class, which serves as a lower bound for performance regarding data imbalance.

  \item \textbf{Voting-Based Confidence Score:} For AutoFL and AutoCodeRover, we utilize the confidence score derived from majority voting as a heuristic baseline. As described in Section~\ref{sec:autofl}, AutoFL~\cite{kang2024autofl} defines confidence as the voting score of the final consensus answer, quantifying the confidence of its inference. We hypothesize that higher confidence correlates with higher correctness. To derive classification baseline, we determine an optimal threshold $\tau \in \{0.1, 0.2, \dots, 0.9\}$ that maximizes classification accuracy on the validation set, and apply it to the test set. Note that the confidence score can only be computed after completing all $R$ trajectories. Therefore, while we adopt it as a baseline for RQ1, this metric cannot be utilized for early termination or hotswapping.
  
  \item \textbf{Lachesis:} For RQ1, which evaluates prediction performance on completed trajectories, we compare our approach with Lachesis~\cite{kim2025lachesis}. Although Lachesis does not support early termination, comparing its correctness prediction results on full graphs allows us to validate the effectiveness of our refined implementation and embedding strategy. We choose the GCN-FAA configuration from the original paper~\cite{kim2025lachesis}, as it was reported to be the best-performing configuration.
\end{itemize}

For evaluation metrics, we employ accuracy, computed at a threshold of 0.5, for overall correctness, AUROC for discriminative capability across thresholds, and AUPR for robustness on imbalanced data. We also report FPR@95 to measure the false alarm rate at 95\% recall. For RQ2 and RQ3, we report monetary cost of running inferences on open weight SLMs, i.e., Llama3, Mixtral, based on the per-token cost of a cloud-based LLM provider\footnote{Together AI: \url{https://www.together.ai/pricing\#serverless-inference}.} and the number of tokens required by inferences. The cost of proprietary models, i.e., GPT-3.5-turbo, GPT-4, and GPT-4o, are reported using their per-token cost\footnote{OpenAI: \url{https://openai.com/api/pricing/}} and the number of tokens required by inferences.

\subsection{Parameters \& Implementations}
We implement \name in Python 3.10. Experiments for the GCN and Llama-3-8B are conducted on a single NVIDIA RTX 3090 GPU, while other models are accessed through cloud-based API services. For the GCN model, we use the Adam optimizer with a learning rate of 0.001, batch size of 32, a hidden dimension of 32, and dropout rate of 0.8. We perform 5-fold cross-validation, splitting the hold-out set of each fold into validation and test subsets. We retain the model checkpoint achieving the best validation performance and conduct the final evaluation on the test set. Consequently, since only half of the hold-out data is used for evaluation in each fold, the total aggregated test set size amounts to 50\% of the full dataset. 

\begin{table*}[ht]
  \caption{Performance comparison of correctness prediction on completed trajectories.}
  \label{tab:rq1_result}
  \centering
  \resizebox{\textwidth}{!}{
  \begin{tabular}{llc|cccc|ccc|ccc|ccc}
    \toprule
    & & & \multicolumn{4}{c|}{\textbf{Accuracy} $\uparrow$} & \multicolumn{3}{c|}{\textbf{AUROC} $\uparrow$} & \multicolumn{3}{c|}{\textbf{AUPR} $\uparrow$} & \multicolumn{3}{c}{\textbf{FPR@95} $\downarrow$} \\
    \textbf{Agent} & \textbf{Model} & \textbf{Criterion} & \textit{Maj.} & \textit{Conf.} & \textit{Lach.} & \textbf{\name} & \textit{Conf.} & \textit{Lach.} & \textbf{\name} & \textit{Conf.} & \textit{Lach.} & \textbf{\name} & \textit{Conf.} & \textit{Lach.} & \textbf{\name} \\ 
    \midrule
    
    \multirow{6}{*}{AutoFL} & \multirow{3}{*}{Llama-3} 
      & Acc@1 & 0.72 & \textbf{0.78} & 0.72 & 0.76 & \textbf{0.83} & 0.60 & 0.73 & \textbf{0.91} & 0.81 & 0.87 & \textbf{0.62} & 0.91 & 0.63 \\
    & & Acc@3 & 0.60 & \textbf{0.75} & 0.63 & 0.68 & \textbf{0.82} & 0.67 & 0.72 & \textbf{0.85} & 0.76 & 0.78 & \textbf{0.63} & 0.84 & 0.67 \\
    & & Acc@5 & 0.56 & \textbf{0.71} & 0.64 & 0.65 & \textbf{0.82} & 0.68 & 0.69 & \textbf{0.83} & 0.72 & 0.70 & \textbf{0.66} & 0.82 & 0.76 \\ \cmidrule{2-16}
    
    & \multirow{3}{*}{GPT-4o} 
      & Acc@1 & 0.56 & 0.70 & 0.72 & \textbf{0.84} & 0.78 & 0.80 & \textbf{0.89} & 0.76 & 0.77 & \textbf{0.90} & 0.68 & 0.60 & \textbf{0.44} \\
    & & Acc@3 & 0.69 & 0.75 & 0.74 & \textbf{0.85} & 0.77 & 0.79 & \textbf{0.90} & 0.65 & 0.65 & \textbf{0.85} & 0.73 & 0.64 & \textbf{0.42} \\
    & & Acc@5 & 0.72 & 0.77 & 0.75 & \textbf{0.86} & 0.75 & 0.76 & \textbf{0.90} & 0.61 & 0.59 & \textbf{0.83} & 0.74 & 0.68 & \textbf{0.37} \\ \midrule
    
    \multirow{6}{*}{AutoCodeRover} & \multirow{3}{*}{Mixtral} 
      & Acc@1 & 0.78 & 0.78 & 0.78 & \textbf{0.88} & 0.79 & 0.75 & \textbf{0.85} & 0.92 & 0.91 & \textbf{0.93} & 0.83 & 0.64 & \textbf{0.34} \\
    & & Acc@3 & 0.72 & 0.73 & 0.75 & \textbf{0.87} & 0.79 & 0.74 & \textbf{0.86} & 0.90 & 0.88 & \textbf{0.92} & 0.83 & 0.69 & \textbf{0.35} \\
    & & Acc@5 & 0.71 & 0.72 & 0.75 & \textbf{0.82} & \textbf{0.80} & 0.78 & \textbf{0.80} & \textbf{0.90} & \textbf{0.90} & 0.89 & 0.83 & 0.61 & \textbf{0.50} \\ \cmidrule{2-16}
    
    & \multirow{3}{*}{GPT-4} 
      & Acc@1 & 0.59 & 0.64 & 0.74 & \textbf{0.93} & 0.68 & 0.82 & \textbf{0.93} & 0.75 & 0.87 & \textbf{0.95} & 0.96 & 0.57 & \textbf{0.17} \\
    & & Acc@3 & 0.58 & 0.62 & 0.59 & \textbf{0.90} & 0.61 & 0.56 & \textbf{0.90} & 0.57 & 0.51 & \textbf{0.91} & 1.00 & 0.81 & \textbf{0.19} \\
    & & Acc@5 & 0.62 & 0.66 & 0.69 & \textbf{0.91} & 0.60 & 0.71 & \textbf{0.90} & 0.53 & 0.63 & \textbf{0.90} & 1.00 & 0.71 & \textbf{0.19} \\ \midrule

    \multirow{3}{*}{RepairAgent} & \multirow{3}{*}{GPT-3.5} 
      & $N_{ppe}\ge 1$ & 0.62 & - & - & \textbf{0.78} & - & - & \textbf{0.85} & - & - & \textbf{0.89} & - & - & \textbf{0.51} \\
    & & $N_{ppe}\ge 3$ & 0.76 & - & - & \textbf{0.81} & - & - & \textbf{0.81} & - & - & \textbf{0.92} & - & - & \textbf{0.52} \\
    & & $N_{ppe}\ge 5$ & 0.80 & - & - & \textbf{0.84} & - & - & \textbf{0.80} & - & - & \textbf{0.93} & - & - & \textbf{0.54} \\ 

    \bottomrule
  \end{tabular}
  }
\end{table*}

\begin{figure} [ht]
    \centering
    \includegraphics[width=1\linewidth]{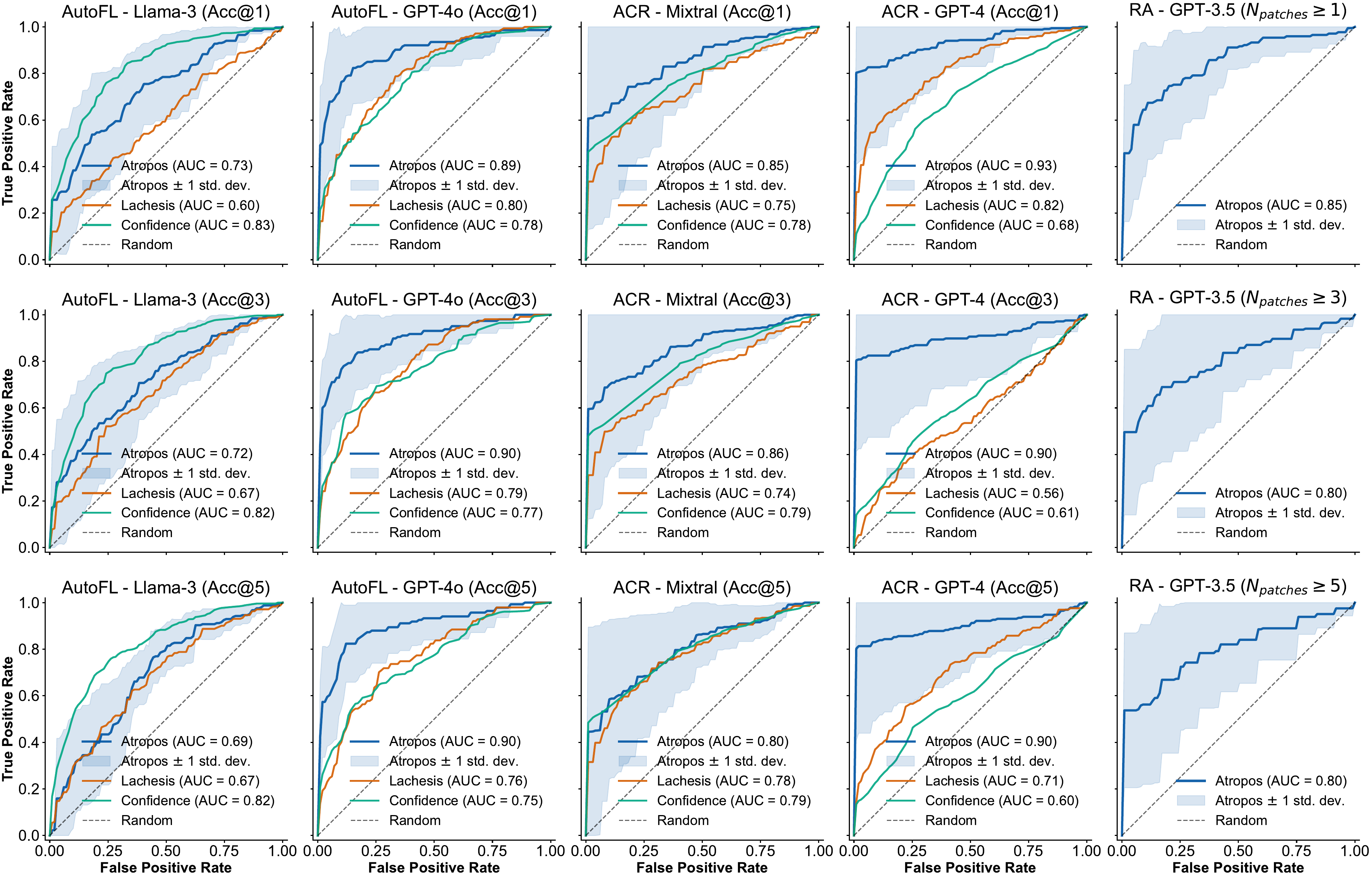}
    \caption{ROC curves of correctness prediction on completed inference across agents and models. The first two columns correspond to AutoFL, followed by two columns for AutoCodeRover, and the final column represents RepairAgent. The rows correspond to \lc{1}, \lc{3}, and \lc{5}, respectively. The shaded regions indicate $\pm1$ standard deviation across cross-validation folds of \name. Note that the AUC values annotated here may slightly differ from those in \Cref{tab:rq1_result} due to the interpolation process applied for visualization.}
    \label{fig2_auroc}
\end{figure}

\section{Results}
\label{sec:results}

Here we report results of our empirical evaluation. 

\subsection{RQ1: Correctness Prediction Accuracy}

\Cref{tab:rq1_result} shows the accuracy of correctness prediction by \name using full trajectories, against different criteria (\lc{1}, \lc{3}, and \lc{5}) for successful solution. For example, when ranking the faulty method at the top is the criterion for success (i.e., \lc{1}), \name can predict whether a given set of 10 inference trajectories taken from GPT-4o is correct or not with an accuracy of 0.84. The best value for each configuration of (agent, model, criterion) is typeset in bold. Note that RepairAgent lacks baselines other than majority class, as its unstructured patch output facilitates neither voting-based confidence scores nor the application of the existing work, Lachesis.

\name outperforms baselines for most configurations. Notably, \name achieves accuracy of 0.93 for AutoCodeRover, run on GPT-4 with \lc{1}, and FPR@95 of 0.17. Further, we note that \name maintains stable performance across all configurations, unlike baselines. The same trend can be observed in \Cref{fig2_auroc}, where the AUROC curves of \name (blue lines) consistently maintain a substantial margin above the random baseline across diverse configurations. 

One exception is the results for AutoFL setup with Llama-3, for which the confidence score yields better results. As Llama-3-8B is the smallest model used in our experiments, it frequently generates abnormal function calls with syntactic errors, resulting in failed reasoning steps. Such behavior is observed consistently regardless of the correctness of the final solution, meaning that the generated SFGs lack sufficient discriminative features to distinguish between correct and incorrect inferences.

On a related point, we also observe that prediction performance is generally higher with trajectories generated by more capable models (GPT-4o, GPT-4) than those by open-weight SLMs. More capable models tend not to generate abnormal function invocations. We hypothesize that higher quality of reasoning steps taken by more capable models leads to more discriminative power in the resulting SFGs.

\name outperforms Lachesis~\cite{kim2025lachesis} in most configurations. This performance gap highlights the advantage of our embedding scheme over the discrete one-hot encoding used previously. For example, consider two reasoning steps involving \verb|NumberUtils| and \verb|NumberUtils.isNumber| respectively. The one-hot encoding of Lachesis would consider these two function arguments as unique, whereas the FastText embeddings used by \name would reflect the similarity between two arguments, which in turn is fed into GCN via node representation, highlighting the semantic connection between two steps.

\subsection{RQ2: Effectiveness of Early Termination}

\begin{figure*}[ht]
    \centering
    \includegraphics[width=0.95\textwidth]{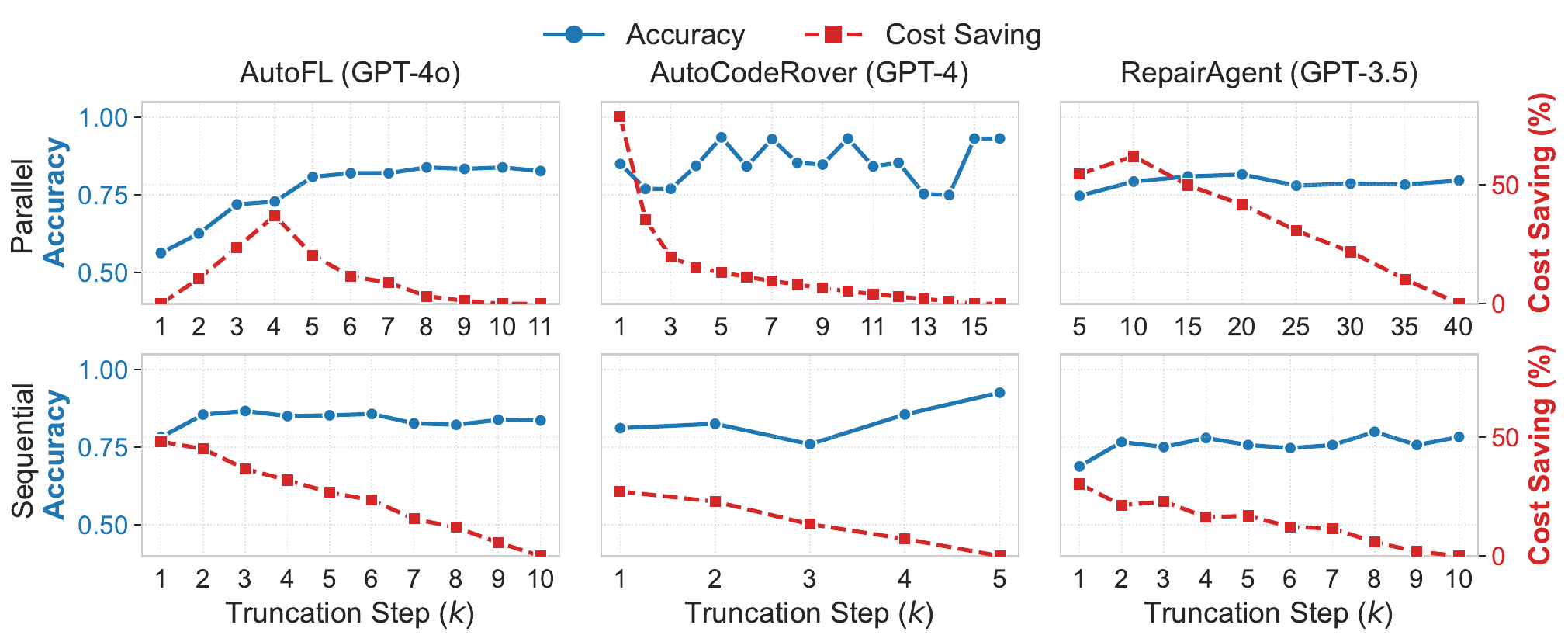}

    \caption{Trade-off analysis between Prediction Accuracy and Cost Saving. The left y-axis (blue) represents prediction accuracy, and the right y-axis (red) indicates the monetary cost saving rate.\label{fig3_tradeoff_accuracy}}
    
\end{figure*}

To answer RQ2, we investigate the trade-offs between prediction accuracy and cost saving rates at different reasoning steps ($k$). \Cref{fig3_tradeoff_accuracy} illustrates such trade-offs for both parallel and sequential approaches, for \lc{1}. We focus on proprietary models, as they offer substantially larger cost-saving opportunities than open-weight SLMs.\footnote{Results from SLMs are available in the artifacts.}  Note that, for ACR, the cost is only for the fault localization step, and excludes the preceding the bug reproduction stage. Similarly, for RepairAgent, we exclude the cost of using auxiliary models that are used to generate mutants for patch validation.

We observe that the overall prediction accuracy remains consistently high across most of the configurations, except for the very early stages of AutoFL in the parallel approach. Specifically, the prediction accuracy for ACR at the midpoint ($k=8$) in the parallel approach is 0.85, with an AUROC of 0.85. This suggests that partial SFGs, representing incomplete inference trajectories, still retain sufficient information regarding the correctness of the final outcome. 

\begin{table}[ht]
    \centering
    \caption{Average number of nodes in SFGs analyzed at different truncation stages. The stages correspond to the fraction of the maximum interaction budget $N \text{ or } N+1$ (for Parallel) or repetitions $R$ (for Sequential).}
    \label{tab:num_nodes}
    \resizebox{\textwidth}{!}{
    \begin{tabular}{ll cccc|ll cccc}
\toprule
\multirow{2.5}{*}{\textbf{Agent}} & \multirow{2.5}{*}{\textbf{Mode}} & \multicolumn{4}{c|}{\textbf{Truncation Progress}} & \multirow{2.5}{*}{\textbf{Agent}} & \multirow{2.5}{*}{\textbf{Mode}} & \multicolumn{4}{c}{\textbf{Truncation Progress}} \\ \cmidrule(lr){3-6}\cmidrule(lr){9-12}
& & \textbf{25\%} & \textbf{50\%} & \textbf{75\%} & \textbf{100\%} & & & \textbf{25\%} & \textbf{50\%} & \textbf{75\%} & \textbf{100\%} \\
\midrule
\multirow{2}{*}{\shortstack[l]{AutoFL\\(GPT-4o)}}       &  Parallel    & 2.38  & 9.33  & 15.27 & 16.99 & \multirow{2}{*}{\shortstack[l]{RepairAgent   \\ (GPT-3.5)}} &  Parallel    & 25.91 & 38.20 & 46.60 & 52.06 \\
                                                        &  Sequential  & 9.71  & 13.61 & 15.21 & 16.99 &                                                             &  Sequential  & 23.35 & 37.31 & 44.07 & 52.06 \\
\midrule
\multirow{2}{*}{\shortstack[l]{AutoCodeRover\\(GPT-4)}} &  Parallel    & 9.16  & 10.16 & 10.27 & 10.30 &                                                             &              &       &       &       & \\
                                                        &  Sequential  & 3.46  & 5.45  & 7.23  & 10.30 &                                                             &              &       &       &       & \\
\bottomrule
    \end{tabular}%
    }
\end{table}

Our analysis of AutoFL results show that, in parallel approach, the lower performance in the early stages is due to the lack of variations in early function call sequences. AutoFL is instructed to begin inferences by calling \verb|get_failing_test_covered_classes|: it tends to follow up with \verb|get_failing_test_covered_method_for_class| using the result of first function call~\cite{kang2024autofl}. \Cref{tab:num_nodes} shows that the average number of nodes in AutoFL SFGs under the parallel setting is only 2.38 at the 25\% progress point ($k=2$): the first two reasoning steps across all 10 trajectories are largely identical, with minimal graph branching. We suspect that the homogeneity in SFGs leads to lack of discriminative features, resulting in low initial accuracy, which in turn impacts the cost saving rate. With other agents, the cost saving is much higher if we check for termination at earlier steps: accurate predictions can save more tokens by terminating incorrect inferences. With AutoFL, the low prediction accuracy in earlier reasoning steps results in lower saving rate, up to step 4. After $k=4$, the cost saving rate begins to decrease like with other agents, once the trajectories diverge and the model acquires sufficient predictive capability. We do note that the low accuracy of AutoFL results in conservative early termination with low false positive: it rarely terminates correct inferences, but allows incorrect inferences to continue at additional cost.

In contrast, AutoCodeRover allows multiple function calls within a single reasoning step, leading to diverse function-call combinations from the very beginning. As shown in \Cref{tab:num_nodes}, ACR exhibits a higher number of unique nodes at the 25\% point ($k=3$) in the parallel setting. Consequently, its graphs contain more information early on, preventing the trend observed with AutoFL. Further, although the maximum interaction length is set to 15 ($N=15$), many ACR trajectories terminate earlier, often within $k=3$. This results in a very high cost saving rate at earlier steps, which subsequently declines rapidly within a few steps. Across all agents, the cost saving rate stabilizes in later stages, as many valid trajectories naturally terminate before reaching high $k$ values, leaving fewer opportunities for early termination.

In the sequential setting, $k$ denotes the number of completed trajectories. Cost savings decrease monotonically as $k$ increases, as the token usage accumulates incrementally with each additional trajectory. Notably, prediction accuracy remains high even at $k=1$, indicating that embedding a single execution trajectory, without the structural information of multiple inferences, still yields significantly accurate predictions. For scenarios where budget is constrained, the sequential approach with $k=1$ presents a viable option, offering maximum cost savings and eliminating the overhead of running the agent multiple times. 

\begin{figure*}[ht]
    \centering
    \includegraphics[width=0.85\textwidth]{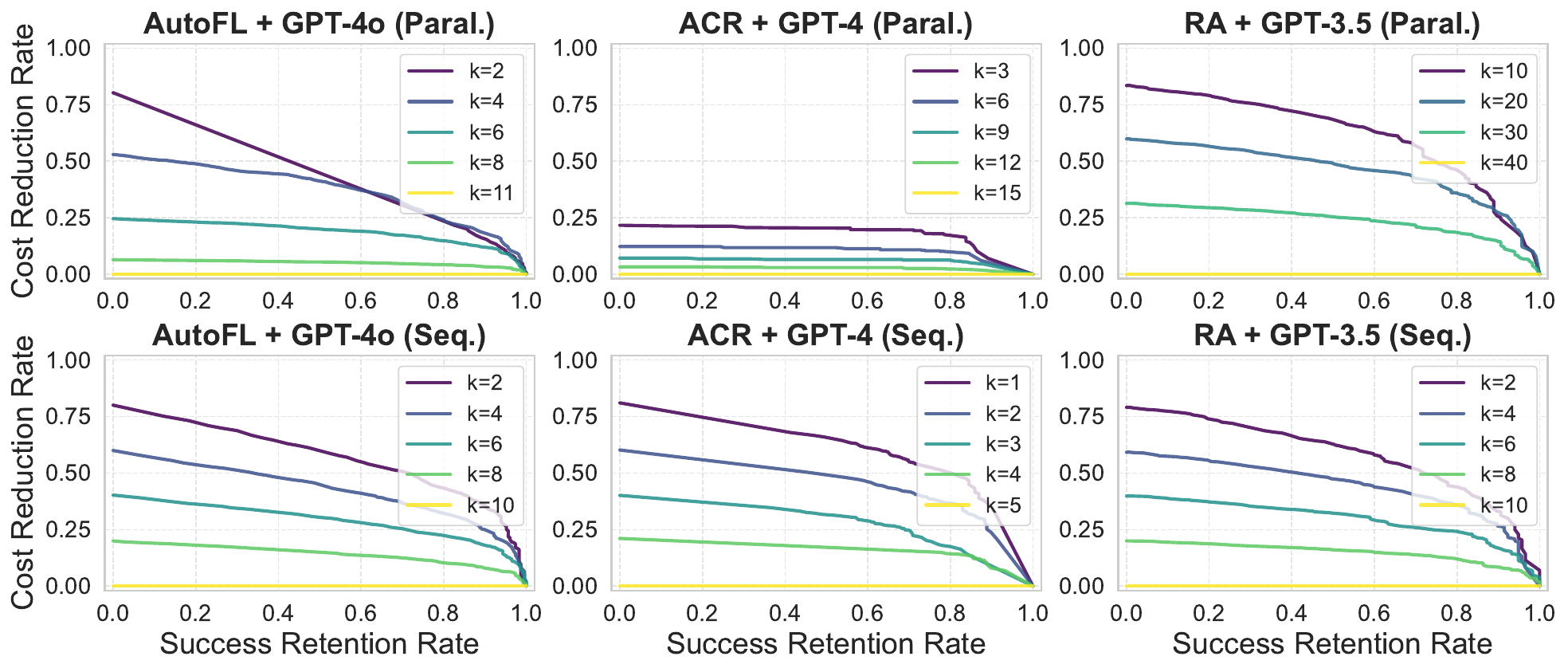}
    \caption{Trade-off analysis between Success Retention Rate and Cost Reduction Rate. The top and bottom rows show the \textit{Parallel} and \textit{Sequential} strategies, respectively.\label{fig4_tradeoff_retention}}
\end{figure*}

\Cref{fig4_tradeoff_retention} illustrates the trade-off between the success retention rate and the cost reduction rate for the results reported in \Cref{fig3_tradeoff_accuracy}. By success retention rate, we mean the fraction of originally successful inferences that remain successful after applying early termination. The objective of early termination is to effectively detect incorrect inferences without prematurely stopping correct ones. Therefore, it is essential to achieve significant cost savings even when the classification threshold is adjusted to maintain a high success retention rate, which would correspond to flat curves in \Cref{fig4_tradeoff_retention} (i.e., higher cost reduction rate at higher success retention rate). The curves in \Cref{fig4_tradeoff_retention} becomes flatter as $k$ increases. This, in turn, provides valuable guidance when selecting $k$ as well as the threshold for early termination for \name. Results in \Cref{fig4_tradeoff_retention} suggest that the point where prediction accuracy begins to saturate would generally be a good choice for the termination threshold, as it would maximize efficiency without compromising the reliability of the correctness prediction.

\subsection{RQ3: Hotswap Effectiveness}

We now turn to the effectiveness of hotswapping based on the early termination prediction of \name. We conduct evaluation of parallel hotswapping at a fixed point of $k=\lceil N/2 \rceil$ for all agents. This constraint is necessitated by the significant computational and financial costs: unlike the sequential approach which can be simulated by combining pre-existing complete trajectories, parallel hotswapping requires the actual re-executions of the agent from step $k$ using the target model given the context of the source model. Consequently, we restricted our evaluation for the parallel approach to this single split point. Further, the performance and the cost of RepairAgent on the target model (GPT-4o) are estimated, as the large interaction budget ($N=40$) and the number of required input/output tokens are significantly higher than other agents. This restricted our evaluation of RepairAgent on the target model (GPT-4o) to a single run of $R=1$. The performance of RepairAgent on the target model is extrapolated based on the ratio between performance under $R=1$ and $R=10$ on the source model (GPT-3.5-turbo); the cost is estimated as ten times that of the cost a single run ($R=1$) under the target model (GPT-4o). However, since we only have a sample size of one ($R=1$), RepairAgent has been excluded from the evaluation of sequential hotswapping.

\Cref{tab:parallel_hotswap_results} summarizes the cost-effectiveness of parallel hotswapping for all three agents. The results show that hotswapping can retain a significant portion of the target model's performance, while incurring only a fraction of the cost. AutoFL exhibits the most efficient trade-off: with hotswapping, it recovers 74.35\% of the performance at acc@1 of GPT-4o, while consuming only 23.90\% of the cost. Similarly, RepairAgent achieves 78.92\% of the target model's performance with 64.73\% of the cost. These results confirm that hotswapping effectively filters out unpromising trajectories early on and selectively invests more resources on those instances that necessitate higher reasoning capabilities, resulting in a better cost-benefit trade-off.

For ACR, the cost reduction is less pronounced compared to other agents, as parallel hotswapping still costs 83.28\% of that of the target model. This is attributed to the rapid problem-solving capability of the target model (GPT-4): in many cases, ACR running on GPT-4 successfully resolves the issue within a few steps. Consequently, the initial steps executed by the source model do not effectively substitute for the target model's workload. Instead, they become sunk costs that add to the total expense without contributing much to the solution. Based on this, we posit that the point of hotswap needs to be carefully tuned for each agent to achieve ideal cost-benefit trade-off.

\begin{table*}[ht]
  \caption{Cost-effectiveness of parallel hotswapping. Comparison of Source-only, Target-only, and Hotswap approaches. Values in parentheses indicate the percentage relative to the Target-only model (set as 100\%). For AutoFL and ACR, \lc{1}, \lc{3}, and \lc{5} correspond to acc@1, acc@3, and acc@5; for RepairAgent, they denote the number of bugs resolved with at least 1, 3, and 5 patches ($N_{ppe} \ge 1$, $N_{ppe} \ge 3$, and $N_{ppe} \ge 5$)}
  \label{tab:parallel_hotswap_results}
  \centering
  \resizebox{\textwidth}{!}{%
  \begin{tabular}{lr l r@{\hspace{5pt}}rrrr}
    \toprule
    Agent & \# Bugs & Method & \multicolumn{2}{c}{Cost (\$) (Ratio)} & \lc{1} (Ratio) & \lc{3} (Ratio) & \lc{5} (Ratio) \\
    \midrule

    & & Source (Llama-3) & 2.63 & (0.91\%) & 120 (52.17\%) & 172 (58.50\%) & 189 (61.36\%) \\
    \rowcolor{gray!10} 
    & & Hotswap & 69.01 & (23.90\%) & 171 (74.35\%) & 239 (81.29\%) & 255 (82.79\%) \\
    \multirow{-3}{*}{AutoFL} & \multirow{-3}{*}{428} 
    & Target (GPT-4o) & 288.80 & (100.00\%) & 230 (100.00\%) & 294 (100.00\%) & 308 (100.00\%) \\ 
    \midrule

    & & Source (Mixtral) & 129.52 & (24.38\%) & 111 (53.88\%) & 132 (50.38\%) & 142 (48.63\%) \\
    \rowcolor{gray!10} 
    & & Hotswap & 442.44 & (83.28\%) & 160 (77.67\%) & 212 (80.92\%) & 239 (81.85\%) \\
    \multirow{-3}{*}{AutoCodeRover} & \multirow{-3}{*}{500} 
    & Target (GPT-4) & 531.29 & (100.00\%) & 206 (100.00\%) & 262 (100.00\%) & 292 (100.00\%) \\ 
    \midrule

    & & Source (GPT-3.5) & 322.90 & (12.93\%) & 117 (63.24\%) & 88 (63.31\%) & 73 (63.48\%) \\
    \rowcolor{gray!10} 
    & & Hotswap & 1616.94 & (64.73\%) & 146 (78.92\%) & 108 (77.70\%) & 89 (77.39\%) \\
    \multirow{-3}{*}{RepairAgent} & \multirow{-3}{*}{305} 
    & Target (GPT-4o) & 2497.82 & (100.00\%) & 185 (100.00\%) & 139 (100.00\%) & 115 (100.00\%) \\ 
    \bottomrule
  \end{tabular}
  }
\end{table*}

Next, we examine the sequential hotswapping strategy. Unlike the parallel approach, which requires a fixed split point due to re-execution costs, the sequential setting enables an evaluation across all $k$ values by combining existing trajectories. In this setting, $k$ represents the number of trajectories fully executed by the source model. If the predictor forecasts failure based on these $k$ completed trajectories, the subsequent $R-k+1$ trajectories, including the $k$-th one, are generated on the target model. RepairAgent is excluded from this evaluation due to reasons described above.

\begin{figure*}[ht]
    \centering
    \includegraphics[width=1\textwidth]{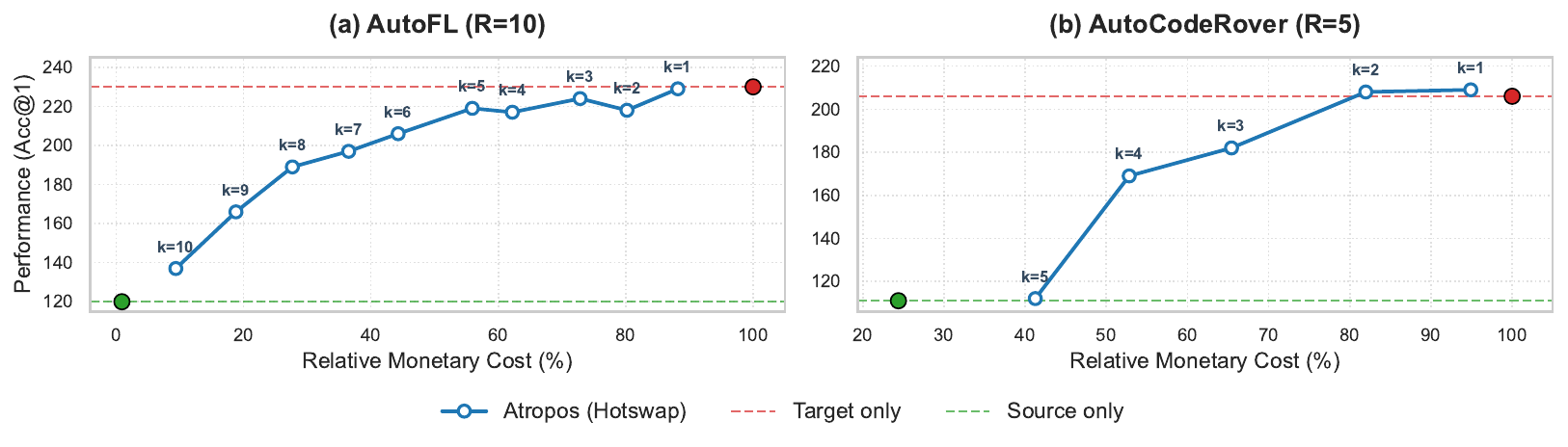}
    \caption{Cost-benefit trade-off of sequential hotswapping. The x-axis represents the cost relative to the Target-only baseline (100\%). The markers on the curve correspond to split points $k$, representing the hotswap point.\label{fig5_sequential_hotswap}}
\end{figure*}

\Cref{fig5_sequential_hotswap} illustrates the cost-benefit trade-offs of the sequential hotswapping. The results for both AutoFL and AutoCodeRover exhibit a convex curve towards the upper-left corner, indicating a highly efficient trade-off. Specifically, for AutoFL (\Cref{fig5_sequential_hotswap} (a)), the localization performance remains stable near the Target-only baseline up to $k=5$. This suggests that replacing half of the expensive target model's executions with the source model does not compromise the final performance.

We note that, with AutoCodeRover, hotswapping at $k=1$ and $k=2$ yields performance that surpasses the Target-only baseline (\Cref{fig5_sequential_hotswap} (b)). We consider this as an unexpected side-effect of model diversity enabled by hotswapping, as reported in the literature with ensembles of LLMs~\cite{Cho2025aa}: the consensus formed by combining outputs from both the source model (Mixtral) and the target model (GPT-4) appears to be more accurate than relying exclusively on the target model. This suggests that the source model, despite being generally weaker, contributes unique correct patches or diverse perspectives that aid the majority voting process. As $k$ increases beyond the midpoint, the voting power of the target model naturally diminishes as fewer trajectories are allocated to it. Consequently, both cost and accuracy gradually converge towards the Source-only baseline. This trend demonstrates that $k$ can be adjusted to balance cost with performance requirements. 

\begin{figure*}[t]
    \centering
    \includegraphics[width=1\textwidth]{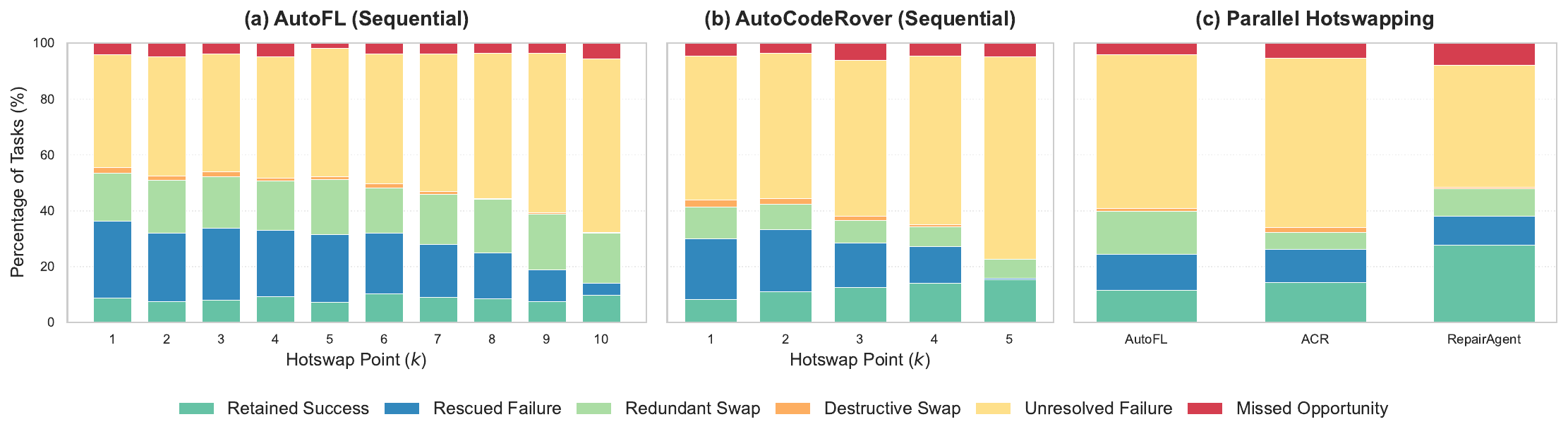}
    \caption{Categorization of final inference outcomes after hotswapping: (a) and (b) illustrate the distribution of outcomes for sequential hotswapping for AutoFL and ACR, respectively. (c) presents the results for parallel hotswapping at the inference midpoint.\label{fig6_sequential_hotswap_breakdown}}
\end{figure*}

We also investigate the efficacy of the hotswap intervention by categorizing final inferences based on the correctness of the source model, the prediction, and the post-hotswap outcome. \textit{Retained Success} refers to cases where a correct source inference is correctly predicted and continued without hotswap. \textit{Rescued Failure} denotes originally failing instances that are correctly predicted and successfully rectified after hotswap. \textit{Unresolved Failure} represents cases where the hotswap fails to fix the inference depite a correct failure prediction. If a correct source trajectory is incorrectly predicted as a failure, it results in either a \textit{Redundant Swap} (if the final outcome is nonetheless successful) or a \textit{Destructive Swap} (if the final outcome is a failure). Finally, \textit{Missed Opportunity} denotes failing trajectories incorrectly predicted as successful and allowed to continue.

\Cref{fig6_sequential_hotswap_breakdown} shows that, for sequential hotswapping, lower $k$ values exhibit a higher proportion of \textit{Rescued Failures} due to the greater contribution of the target model. Specifically, for AutoFL, the proportion of \emph{Rescued Failure} reaches 27.57\% at the step $k=1$. The performance of each agent corresponds to the height of the bottom three parts (retained successes, rescued failures, and redundant swaps): it remains stable up to the midpoint of $k$, indicating that the sequential hotswap point can be flexibly adjusted within this range. We also note that the proportion of \textit{Destructive Swaps} is negligible. This is attributed not only to the precision of the early termination prediction but also to the robustness of the target model, which rarely degrades a correct trajectory, even when the hotswap is unnecessarily invoked. Furthermore, \textit{Missed Opportunities} are minimal, demonstrating the high recall of the early termination prediction in detecting potential failures. Despite the sunk costs incurred by \textit{Unresolved Failures}, the overall analysis in \Cref{tab:parallel_hotswap_results} and \Cref{fig5_sequential_hotswap} confirms that hotswapping can be cost-effective.

\begin{table}[ht]
  \caption{Impact of graph components. Performance comparison (Accuracy and AUROC) of the full model against ablated variants using trajectories generated by target models.}
  \label{tab:ablation}
  \centering
  \resizebox{0.6\columnwidth}{!}{%
  \begin{tabular}{l rr rr}
    \toprule
    \multirow{2}{*}{Variant} & \multicolumn{2}{c}{AutoFL (GPT-4o)} & \multicolumn{2}{c}{AutoCodeRover (GPT-4)} \\
    \cmidrule(lr){2-3} \cmidrule(lr){4-5}
     & Acc. & AUROC & Acc. & AUROC \\
    \midrule
    \name (Full Model) & 0.84 & 0.89 & 0.93 & 0.93 \\
    \midrule
    w/o Semantics (Lachesis) & 0.72 & 0.80 & 0.74 & 0.82 \\
    w/o Argument (Func. Only) & 0.70 & 0.78 & 0.60 & 0.63 \\
    w/o Function (Arg. Only) & 0.81 & 0.88 & 0.63 & 0.70 \\
    \bottomrule
  \end{tabular}
  }
\end{table}

\subsection{RQ4: Ablation Study}

\Cref{tab:ablation} summarizes the performance of each ablation variant for AutoFL and ACR. We compare the full early termination prediction model against three variants: \textit{w/o Semantics}, which replaces semantic embeddings with anonymized n-hot vectors (identical to the FAA setting in Lachesis~\cite{kim2025lachesis}); \textit{w/o Arguments}, which removes argument vectors to rely solely on function types; and \textit{w/o Functions}, which conversely uses only argument embeddings. We perform ablation study using target models only (GPT-4o for AutoFL, GPT-4 for AutoCodeRover), as we expect their greater performance would reveal relative contributions from individual components more clearly.

The significant performance drop in \textit{w/o Semantics} confirms that semantic information contributes significantly to predictive accuracy. Moreover, the clear performance gap between \textit{w/o Semantics} and \textit{w/o Arguments} indicates that the specific targets of agent actions contain essential information. Notably, the contribution of the function type component varies between the two agents. AutoFL experiences only a minor performance drop, implying that the arguments (i.e., classes and methods) contain more information than the action itself. In contrast, ACR exhibits a substantial degradation in performance, implying that its use of a more diverse range of retrieval and navigation actions requires both function types and their arguments for accurate prediction. All in all, these results show that all components in SFG representation are essential for achieving the best performance.

\section{Threats to Validity}
\label{sec:threats_to_validity}

Internal validity concerns whether our observation is truly caused by the proposed technique. The early termination prediction based on SFG of agent inferences does perform better than random prediction as well as the naive baseline, and the result of the ablation study supports that information in SFGs contributes to the prediction accuracy. Hotswapping keeps all factors other than the underlying LLM the same, ensuring that the observed changes in agent performance are due to the swap and not due to changes in tool execution or any other environmental changes. One specific potential threat to internal validity is data leakage from node embeddings: there is a risk that the predictive model relies on specific tokens associated with the correctness of inferences during training, rather than learning generalizable structural patterns of agent inferences. However, our ablation study shows that \name retains predictive power even when semantic information is removed under \emph{w/o Semantic variant} configuration, suggesting that SFGs can capture structural patterns that are indicative of inference correctness. 

External validity concerns factors that may limit the degree to which \name generalizes. While our empirical evaluation includes three different agents with different sets of tools, only broader future studies can mitigate this threat. Construct validity can be threatened if our chosen metrics do not fully reflect what we intend to measure. To measure the accuracy of early termination prediction, we rely on standard evaluation for predictive models. To measure performance after hotswapping, we adopt the metrics originally used by the respective agents to minimize the threat.

\section{Related Work}
\label{sec:related_work}

Several approaches have been proposed to reduce the overall cost of LLM inferences. Model cascading attempts to reduce cost via a sequential invocation strategy, initiating requests with smaller models first and only transitioning to more capable ones when necessary, i.e., when the initial response is incorrect or lacks certainty~\cite{chen2024frugalgpt}. In contrast, model routing~\cite{ong2025routellm, ding2024hybrid} estimates task difficulty and routes queries to models expected most likely to be capable of resolving them, based on the predicted task complexity. However, these approaches predominantly focus on single-turn question answering, or relatively simple generation tasks. Unlike existing work, \name targets more complex agentic tasks that involve multi-step reasoning and external tool invocations. Further, while routing and cascading typically rely on the analysis of either the initial input or the final output of an inference, \name leverages intermediate structural properties of ongoing agent executions. By analyzing SFGs of partial trajectories, \name can predict eventual failures early on and trigger intervention, preventing any unnecessary computation on the source model from being wasted. The prediction of correctness in the middle of inferences is also different from uncertainty estimations that saves labeling cost~\cite{chen2024inside,SuEtAl2024unsupervised,HeEtAl2024LLM}, as these techniques require inferences to finish.

\section{Conclusion}
\label{sec:conclusion}

We present \name, a technique that aims to improve the cost-benefit trade-off of LLM-based agents using early termination prediction and model hotswapping. Early termination can save computational costs that would have been wasted for unsuccessful inferences; model hotswapping can salvage such inferences by migrating inference contexts from a weak SLM to a more capable LLM and continuing them. Using both techniques, \name can achieve 74.35\% of the performance of a stronger LLM using only 23.9\% of the monetary cost. Future work will consider extending the predictive modelling of \name to incorporate the latent space of source LLMs to improve accuracy.

\section*{Data Availability}
All source code and datasets, including raw execution logs of agents, are available at \url{https://anonymous.4open.science/r/atropos-4683}.

\bibliographystyle{ACM-Reference-Format-num}
\bibliography{ref}
\end{document}